\documentclass[acmsmall, screen]{acmart}
\setcopyright{none}

\acmDOI{10.1145/3797968}

\acmJournal{TQC}
\acmVolume{7}
\acmNumber{2}
\acmArticle{14}
\acmMonth{4}

\usepackage[utf8]{inputenc}
\usepackage{siunitx}
\usepackage{multirow}
\usepackage{subcaption}

\newcommand{\us}[1]{\SI{#1}{\micro\second}}
\newcommand{\um}[1]{\SI{#1}{\micro\meter}}
\newcommand{\SEC}{\text{SEC}}
\newcommand{\SC}{\text{SC}}
\newcommand{\atom}{\text{atom}}
\newcommand{\CZ}{\operatorname{CZ}}

\begin{document}

\title{Benchmarking fault-tolerant quantum computing hardware via QLOPS}

\author{Linghang Kong}
\orcid{0000-0002-5854-5340}
\author{Fang Zhang}
\orcid{0000-0002-0000-7101}
\affiliation{%
  \institution{Zhongguancun Laboratory}
  \city{Beijing}
  \country{P. R. China}
}

\author{Jianxin Chen}
\affiliation{%
  \institution{Tsinghua University}
  \city{Beijing}
  \country{P. R. China}}
\email{chenjianxin@tsinghua.edu.cn}
\orcid{0000-0002-9365-776X}


\begin{abstract}
It is widely recognized that quantum computing has profound impacts on multiple fields, including but not limited to cryptography, machine learning, materials science, etc. To run quantum algorithms, it is essential to develop scalable quantum hardware with low noise levels and to design efficient fault-tolerant quantum computing (FTQC) schemes. Currently, various FTQC schemes have been developed for different hardware platforms. However, a comprehensive framework for the analysis and evaluation of these schemes is still lacking. In this work, we propose Quantum Logical Operations Per Second (QLOPS) as a metric for assessing the performance of FTQC schemes on quantum hardware platforms. This benchmarking framework will integrate essential relevant factors, e.g., the code rates of quantum error-correcting codes, the accuracy, throughput, and latency of the decoder.  Through a resource analysis of factoring RSA-2048, we demonstrate that QLOPS reflects the practical requirements of quantum algorithm execution. This framework will enable the identification of bottlenecks in quantum hardware, providing potential directions for their development. Moreover, our results will help establish a comparative framework for evaluating FTQC designs. As this benchmarking approach considers practical applications, it may assist in estimating the hardware resources needed to implement quantum algorithms and offers preliminary insights into potential timelines.
\end{abstract}




\maketitle

\section{Introduction}

The field of quantum computing has achieved remarkable progress in the past few decades, fully demonstrating its transformative potential. For example, Shor's algorithm \cite{shor1994algorithms} can efficiently solve group-theoretic problems like integer factoring and discrete logarithm, thereby posing a significant threat to information security by compromising mainstream public-key encryption systems; various algorithms for quantum simulation \cite{pnas.1801723115} hold great significance for materials science; quantum linear algebra algorithms, such as the Harrow-Hassidim-Lloyd (HHL) algorithm \cite{PhysRevLett.103.150502}, can provide quantum speedup for machine learning applications.

However, significant progress still has to be made to quantum hardware for such applications to be implemented. The mainstream quantum hardware platforms include superconducting qubits \cite{huang2020superconducting,10.1063/1.5089550}, neutral atoms \cite{wintersperger2023neutral,bluvstein2024logical} and trapped ions \cite{10.1063/1.5088164}. Breakthroughs have been made in the past few years on these platforms, enabling experimental demonstration of devices with hundreds of qubits with an error rate on the order of $0.1\%$. It remains challenging to reduce the error rate by several orders of magnitude to enable reliable quantum computation directly on quantum hardware. As a result, fault-tolerant quantum computing schemes have to be introduced, which combine multiple physical qubits to encode fewer logical qubits so that the error rate can be significantly reduced. Quantum algorithms are then executed through logical operations on these encoded qubits to produce reliable computational results. This approach poses formidable challenges for the quantum decoding architecture.

Current discussions on fault-tolerant quantum computing primarily focus on the quantum memory experiment, which involves initializing the logical qubits into a predefined state (usually the $|0\ldots0\rangle$ state), repeatedly running the syndrome extraction circuit, and ultimately evaluating if the logical qubits remain in the initial state given the decoding results from the decoder. No logical operation is involved in this analysis. On the one hand, the quantum memory experiment has proven valuable for analyzing key parameters such as decoder throughput, accuracy, and latency. On the other hand, it falls short of addressing the costs associated with logical operations.

There have been proposals on benchmarking the performance of quantum devices across platforms, for example, quantum volume (QV) \cite{PhysRevA.100.032328,8936946,Jurcevic_2021} and its generalizations \cite{BlumeKohout2020volumetricframework,proctor2022measuring}, as well as Circuit Layer Operations Per Second (CLOPS) \cite{wack2021quality}. These benchmarks were defined based on the efficiency of executing a specific type of circuits, and therefore cannot reflect the performance for general quantum algorithms. Furthermore, they were defined for running the circuit directly on the quantum hardware, and did not take into account fault tolerance. Another proposal was reliable
Quantum Operations Per Second (rQOPS) \cite{10.1145/3624062.3624211}, which was defined on fault-tolerant quantum computing hardware, but failed to capture many essential factors introduced by classical resources, such as decoder throughput and latency.

In this work, we benchmark the computational capability of quantum hardware from the perspective of logical operations. We establish Quantum Logical Operations Per Second (QLOPS) as a comprehensive metric that incorporates a wide range of relevant factors, and then calculate the QLOPS of superconducting qubits with the surface code and neutral atom qubits with the generalized bicycle (GB) codes \cite{viszlai2023matching} as an example. Note that these calculations only serve as a demonstration, and that the QLOPS value could be optimized with more careful choices of parameters on these platforms. The implications of this work include the following.
\begin{itemize}
    \item \emph{A holistic evaluation framework.} This metric integrates various factors relevant to fault-tolerant quantum computing and allows a more realistic estimation of progress in the field. Early research often overlooks the cost introduced by classical computation, but as quantum hardware scales, the decoder throughput has emerged as a bottleneck. While a parallel decoder \cite{PRXQuantum.4.040344, skoric2023parallel} can partially solve the throughput problem, limitations such as latency and transmission bandwidth remain significant constraints on hardware performance.
    Existing frameworks often fail to systematically analyze how these interconnected factors collectively impact quantum hardware, so in this work, we aim to provide a unified metric that accounts for these factors.
    \item \emph{Bottleneck identifications for hardware optimizations.} By analyzing how specific hardware parameters influence overall performance, this metric will pinpoint bottlenecks and guide iterative hardware development. For example, estimating a system’s fault tolerance performance under achievable near-term improvements -- such as lower two-qubit gate error rates or longer coherence times -- will quantify the relative impact of parameter enhancements. This analysis will enable hardware teams to prioritize efforts on parameters that offer the highest improvement, accelerating progress toward practical fault tolerance.
    \item \emph{Application-driven scheme comparison.} The metric we propose takes into account the fault-tolerant scheme used in the system, and therefore can be used to compare the schemes on a given quantum hardware. For instance, on the neutral atom platform, qubits could be encoded using quantum low-density parity check (LDPC) codes \cite{viszlai2023matching}, and multiple strategies exist for logical operations: one can convert the logical qubits into surface code and apply lattice surgery \cite{Litinski2019gameofsurfacecodes}, or do generalized surgery \cite{sciadv.abn1717,cowtan2025parallel} directly on the LDPC codes. By analyzing the performance of the same neutral atom hardware equipped with different schemes, one can make a comparison of these schemes.
\end{itemize}

The rest of this work is structured as follows. In Section~\ref{sec:def}, we will first define QLOPS (quantum logical operations per second) as a benchmark of the computational power of fault-tolerant quantum computing hardware. Then in Section~\ref{sec:compare} we evaluate it in some example systems. We start with a simplified model that only contains the encoded computational qubits, and analyze the resources associated with magic-state distillation in the end. The comparison between the superconducting platform and the neutral atom platform can be found in Table~\ref{tab:qlops} and Fig.~\ref{fig:qlops}. We also analyze the systems in the recent RSA2048 resource estimations \cite{gidney2025factor,zhou2025resource}, and see how these application-specific results compare with our general-purpose QLOPS benchmark. We conclude the paper with discussion in Section~\ref{sec:discussion}.

\section{The definition of QLOPS}
\label{sec:def}
Traditionally, the computational capability of a classical processor can be quantified using floating-point operations per second (FLOPS). FLOPS is defined with respect to some precision (e.g. 32-bit floats or 64-bit floats), and is given by the product of three factors:
\begin{enumerate}
    \item The number of cores in the processor;
    \item The clock frequency, i.e., the number of clock cycles per second;
    \item The number of floating-point operations per cycle, which depends on the number of cycles needed for each instruction and the number of parallel operations carried out in a single instruction. This is dictated by the architecture of the processor.
\end{enumerate}

Likewise, fault-tolerant quantum computing hardware could be benchmarked by the number of quantum logical operations per second (QLOPS). Following the definition of FLOPS, we define the QLOPS of a device as the product of the following factors:
\begin{enumerate}
    \item $f_1$: The total number of logical qubits in the device;
    \item $f_2$: The syndrome extraction cycle (SEC) frequency;
    \item $f_3$: The number of logical operations per SEC, which is the inverse of the number of SECs needed per logical operation.
\end{enumerate}

For a more accurate definition, one might want to replace the number of logical qubits by the max number of parallel logical operations that can be applied to the logical qubits. This can be limited by multiple factors, like the rate with which the magic states are produced (assuming a scheme based on magic state distillation, but QLOPS can be studied on general fault-tolerance schemes) and the number of logical qubits simultaneously accessible in the encoding. Here, we work with the simplified model and only consider the total number of logical qubits.

Similarly to FLOPS, QLOPS should be defined with respect to some precision parameter, which is the logical error rate per logical qubit per syndrome extraction cycle. This can be estimated numerically with a quantum memory experiment. Given an $[[n, k, d]]$ code, we first prepare the qubits in a logical $|0\rangle^{\otimes k}$ state, and then perform $d$ layers of syndrome extraction circuits under some circuit-level noise model. The total logical error rate $p_L$ is the probability that the final logical state is different from $|0\rangle^{\otimes k}$ after decoding. The logical error per layer per logical qubit is then defined as 

\begin{equation}
    p_0=1-(1-p_L)^{1/(kd)}. \label{eq:p0}
\end{equation}
Technically, the errors on the $k$ logical qubits are not independent, so $p_0$ only describes some average behavior and might not correspond to a specific logical qubit. But $p_0$ can still serve as an indicator when comparing error rates on code patches containing different number of logical qubits. For example, for two types of code patches with $k_1$ and $k_2$ logical qubits, respectively, a fair comparison would be the combined error rate of $k_2$ patches of the first type and $k_1$ patches of the second type, and this will be reduced to the comparison of $p_0$ of each type.

For a device that encodes a single patch of $[[n, k, d]]$ code, we have $f_1=k$. Let $t_{\SEC}$ be the length of the SEC, so $f_2=1/t_\SEC$. $f_3$ is the inverse of the number of SECs needed per logical operation, which is closely related to the \emph{reaction time}. In quantum error correction, a decoder is needed to decide whether a logical correction is needed after the qubits undergo some physical errors, and the next logical operation depends on such decoding results. Therefore, logical operations are slowed down in a system with a long reaction time. The reaction time describes the time from the end of the relevant SEC to obtaining the decoding result. It includes the computation time of the decoder, and for some systems (e.g., the neutral atom platform in Section~\ref{subsec:atom}), the measurement time is also included, as the ancilla measurements are pipelined with the gates in the next round of SEC. In most fault-tolerant logical operation schemes, $d$ rounds of syndrome extraction are needed so that the time-like distance matches the space-like distance of the code, i.e., the logical error rate caused by measurement errors is of the same order as that caused by data qubit errors. The $d$ rounds of syndrome data will be processed by the decoder, which returns the decoding result needed for the logical operation. Let $t_r$ be the reaction time, so the number of syndrome extraction cycles needed for a logical operation would be $\lceil t_r/t_{\SEC}\rceil+d$. Then the QLOPS of this device is given by

\begin{equation}
    Q=k\times \frac{1}{(\lceil t_r/t_{\SEC}\rceil+d)t_{\SEC}}. \label{eq:qlops}
\end{equation}

Note that such $d$ rounds of syndrome extraction might not be necessary in some carefully designed schemes \cite{cain2024correlated,serraperalta2025decodingtransversalcliffordgates,cain2025fast,turner2025scalable,zhou2025resource}, and an adapted formula should be used when evaluating the QLOPS of a system that employs such schemes. 

Additionally, when $t_r > d\times t_\SEC$, a parallel window decoder \cite{PRXQuantum.4.040344, skoric2023parallel} is needed to avoid exponential delay of the logical operations \cite{RevModPhys.87.307}. Such decoders can introduce additional overhead due to the buffers on each window, as well as multithreading if the decoder runs on a multicore CPU. Here, we work in a simplified setting and ignore such overhead.

QLOPS is defined to analyze specific quantum devices (e.g., the Willow chip \cite{acharya2024quantum} with the $d=7$ surface code), but we can also use it to study a general hardware platform (e.g., superconducting qubits with the surface code). Since taking $m$ copies of any system results in a larger system with $m$ times the QLOPS, it might be helpful to consider the \emph{QLOPS density}, defined as QLOPS divided by the system size, as a benchmark of the efficiency of the physical platform and the fault-tolerant quantum computing scheme. For the analysis of the superconducting and neutral atom platforms below, the ``system size'' refers to the number of physical qubits in the quantum error correction code, including both the data qubits and the ancilla qubits. We assume that the measurements in the neutral atom platform are non-destructive so that ``the number of ancilla qubits'' is meaningful, but the analysis also applies to systems in which the measurement operation is destructive and a constant supply of fresh ancilla qubits is needed for each round.

\section{Comparison of superconducting and neutral atom systems}
\label{sec:compare}
In this section, we evaluate and compare the QLOPS of the superconducting and neutral atom systems, encoded with the surface code and the generalized bicycle (GB) code \cite{viszlai2023matching}, respectively. For simplicity we first focus on the encoded computational qubits, ignoring the auxiliary qubits used for routing and magic state distillation. The cost of magic state distillation will be discussed in Section~\ref{subsec:distillation}.

To make a fair comparison, the systems are analyzed under the same level of overall logical error rates. We will adjust the distance of the surface code on the superconducting platform to match the error rate of the neutral atom systems.

Our analysis serves as a demonstration of QLOPS as a benchmark of fault-tolerant schemes in various quantum hardware platforms, and we leave as future work the design of schemes that optimize QLOPS.

\subsection{Superconducting qubits with the surface code}

We first run the surface code simulation with the parameters of Google's Willow quantum processor \cite{acharya2024quantum} and the expected parameters in ten years shown in Table~\ref{tab:sc-param}. The preparation time and error rate of Willow were not presented in the paper, so we use estimations based on current hardware.

\begin{table}[ht]
    \centering
    \caption{Parameters of superconducting quantum hardware, both the currently available parameters based on the Willow processor and the expected parameters in 10 years.}
    \label{tab:sc-param}
    \begin{tabular}{c|c|c}
        \hline
        Parameters & Current & Future \\
        \hline
        Coherence time & \us{80} & \us{1000} \\
        \hline
        Gate time (1Q) & \us{0.025} & \us{0.02}\\
        \hline
        Gate time (2Q) & \us{0.04} & \us{0.03}\\
        \hline
        Gate infidelity (1Q) & 5e-4 & 1e-4\\
        \hline
        Gate infidelity (2Q) & 2e-3 & 5e-4\\
        \hline
        Readout time & \us{0.5} & \us{0.1}\\
        \hline
        Readout error & 7e-3 & 2e-3\\
        \hline
        Preparation time & \us{0.1} & \us{0.1} \\
        \hline
        Preparation error & 5e-3 & 1e-3\\
        \hline
    \end{tabular}
\end{table}

\begin{figure}[ht]
  \centering
  \includegraphics[width=0.45\textwidth]{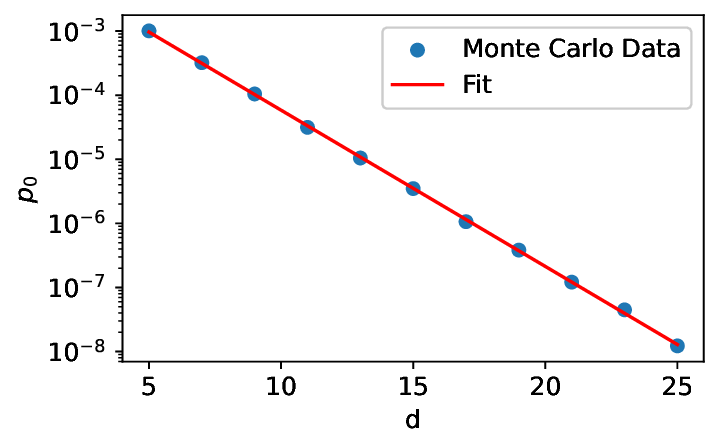}
  \caption{Exponential fit of logical error rate vs code distance}
  \Description{Exponential fit of logical error rate vs code distance}
  \label{fig:exp_fit}
\end{figure}

The syndrome extraction circuit is the one with $\CZ$ gates presented in \cite{McEwen2023relaxinghardware}, and the cycle length is given by the sum of the time needed for state preparation, measurement, 4 single-qubit gates, and 4 $\CZ$ gates, which equals to 0.86 us for the current parameters and 0.40 us for the future parameters. We found that the logical error rate can be fitted to an exponential decay curve. For example, Fig.~\ref{fig:exp_fit} shows the fitting for the logical error rate corresponding to the current parameters. (Technically, we do a linear fitting of $\log(p_0)$ instead of an exponential fitting of $p_0$, because the data is collected with the number of errors fixed, so that the uncertainty is multiplicative.)

The decoding time of PyMatching on the surface code is given in Table~\ref{tab:sc-decoding}. The decoding test was performed on a dual-socket \emph{Intel Xeon Gold 6248R CPU}, featuring a base frequency of 3.00 GHz and a maximum frequency of 4.00 GHz.

\begin{table}[ht]
    \centering
    \caption{Decoding time of the surface code, with \ref{subtab:current} for current parameters and \ref{subtab:future} for future parameters. We only list the data within the range relevant to our calculations.}
    \label{tab:sc-decoding}
    \begin{subtable}[t]{0.5\textwidth}
    \centering
    \caption{}
        \begin{tabular}{c|c|c|c}
            \hline
            $d$ & 11 & 13 & 15 \\
            \hline
            $t_r(\si{\second})$ & 1.2443e-05 & 2.0780e-05 & 3.2834e-05 \\
            \hline
            $d$ & 17 & 19 & 21 \\
            \hline
            $t_r(\si{\second})$ & 9.7008e-05 & 4.8795e-05 & 6.9903e-05  \\
            \hline
            $d$ & 23 & 25 & 27 \\
            \hline
            $t_r(\si{\second})$ & 1.3085e-04 & 1.7153e-04 & 2.2188e-04\\
            \hline
        \end{tabular}
        \label{subtab:current}
    \end{subtable}
    \begin{subtable}[t]{0.5\textwidth}
    \centering
    \caption{}
        \begin{tabular}{c|c|c|c}
            \hline
            $d$ & 5 & 7 & 9 \\
            \hline
            $t_r(\si{\second})$ & 2.0955e-07 & 5.5807e-07 & 1.1682e-06 \\
            \hline
            $d$ & 11 & 13 & 15 \\
            \hline
            $t_r(\si{\second})$ & 2.1191e-06 & 3.4904e-06 & 5.4921e-06  \\
            \hline
        \end{tabular}
        \label{subtab:future}
    \end{subtable}
\end{table}

\subsection{Neutral atoms with generalized bicycle code}
\label{subsec:atom}
For the neutral atom platform, we encode the qubits with GB codes \cite{viszlai2023matching}. GB codes are defined by placing the data qubits and ancilla qubits on a 2-dimensional grid, and each ancilla qubit is associated with a stabilizer check that contains data qubits with fixed offset relative to the ancilla qubit. Then the ancilla qubits could be placed in Acousto-Optic Deflectors (AODs) and moved to their target data qubit together in each step. The time required to move a distance of $\Delta x$ is given by $\sqrt{6\Delta x/a_p}$ \cite{xu2024constant}. We assume that the lattice spacing is \um{5} and $a_p = 0.02\si{\micro\meter}/\si{\micro\second}^2$.

This syndrome extraction cycle in our GB code simulation is slightly different from that described in \cite{viszlai2023matching}, as we take into account the following realistic constraints in hardware. Measurement is carried out on all of the qubits in a region, and there should not be any qubits in this region that are not measured. Therefore, in our simulations the $X$ and $Z$ ancilla qubits are initially placed outside the region where the data qubits are placed and will be moved back to this place before the measurement. We enumerate all possible locations to minimize the routing cost. We assume that the preparation and measurement of ancilla are carefully timed, so that $X$ ancilla qubits are initialized and ready to be moved to their locations when the $Z$ ancilla qubits are about to be moved to their measurement region and vice versa, so that the preparation and measurement of ancilla are in parallel with other parts of the circuit and do not increase $t_\SEC$.

Then the GB code decoding was simulated under a realistic noise model shown in Table~\ref{tab:na-param}. The ``unintended error'' describes the phenomenon that when the Rydberg laser is applied to a region to apply $\CZ$ to pairs of qubits, some error will be introduced to isolated qubits to which $\CZ$ is not applied. ``Movement error'' will be applied to a group of qubits being shuttled due to heating, in addition to the idling error. We first tried the noise parameters achieved in current experiments \cite{bluvstein2024logical}, but it was above the error correction threshold. Then we tried a model with a lower physical error rate, which is a prediction of the system for the next few years.

\begin{table}[ht]
    \centering
    \caption{Parameters of current and future neutral atom hardware.}
    \label{tab:na-param}
    \begin{tabular}{c|c|c}
        \hline
        Pameters & Current & Future \\
        \hline
        Coherence time & \SI{1.5}{\second} & \SI{20}{\second} \\
        \hline
        Gate time (1Q) & \us{0.5} & \us{0.5} \\
        \hline
        Gate time (2Q) & \us{0.2} & \us{0.2} \\
        \hline
        Gate infidelity (1Q) & 1e-3 & 1e-4 \\
        \hline
        Gate infidelity (2Q) & 5e-3 & 1e-3 \\
        \hline
        Unintended error & 1e-3 & 2e-4 \\
        \hline
        Readout time & \us{500} & \us{50} \\
        \hline
        Readout error & 2e-3 & 2e-4 \\
        \hline
        Preparation error & 7e-3 & 2e-4 \\
        \hline
        Movement error & 1e-3 & 1e-4 \\
        \hline
    \end{tabular}
\end{table}

GB codes are decoded with the BP-LSD decoder \cite{hillmann2024localized}. We use the min-sum algorithm for BP with $10^5$ iterations and ``combination sweep'' strategy (\verb|lsd_cs|) with order 10 for LSD. The results are shown in Table~\ref{tab:performance}. Note that, as a correlated decoder, BP-LSD can decode with improved accuracy if both $X$ syndromes and $Z$ syndromes are given as input. In Table~\ref{tab:performance}, ``Z'' means decoding with only $Z$ syndromes and ``ALL'' means decoding with all of ``X'' and ``Z'' syndromes. We find that decoding with all the syndromes can indeed improve the accuracy, but the decoding time is significantly increased.

\begin{table*}[ht]
\centering
\caption{Logical error rate, decoding time and syndrome extraction cycle length of the codes.}
\label{tab:performance}
\begin{tabular}{c|c|c|c}
\hline
Code & {$p_L$} & {$t_r$ (\si{\second})} & {$t_{\SEC}$ (\si{\second})} \\
\hline
[[72, 12, 6]] Z   & 0.0008372 & 0.000633  & 0.002677 \\
\hline
[[90, 8, 10]] Z   & 0.0001614 & 0.002070  & 0.002911 \\
\hline
[[108, 8, 10]] Z  & 7.40e-05  & 0.002194  & 0.002799 \\
\hline
[[144, 12, 12]] Z & 7.10e-05  & 0.004644  & 0.002901 \\
\hline
[[288, 12, 18]] Z & 1.198e-06 & 0.028281  & 0.003550 \\
\hline
[[72, 12, 6]] ALL & 0.0003286 & 0.066876  & 0.002677 \\
\hline
\end{tabular}
\end{table*}

\begin{figure}[ht]
    \centering
    \includegraphics[width=0.45\textwidth]{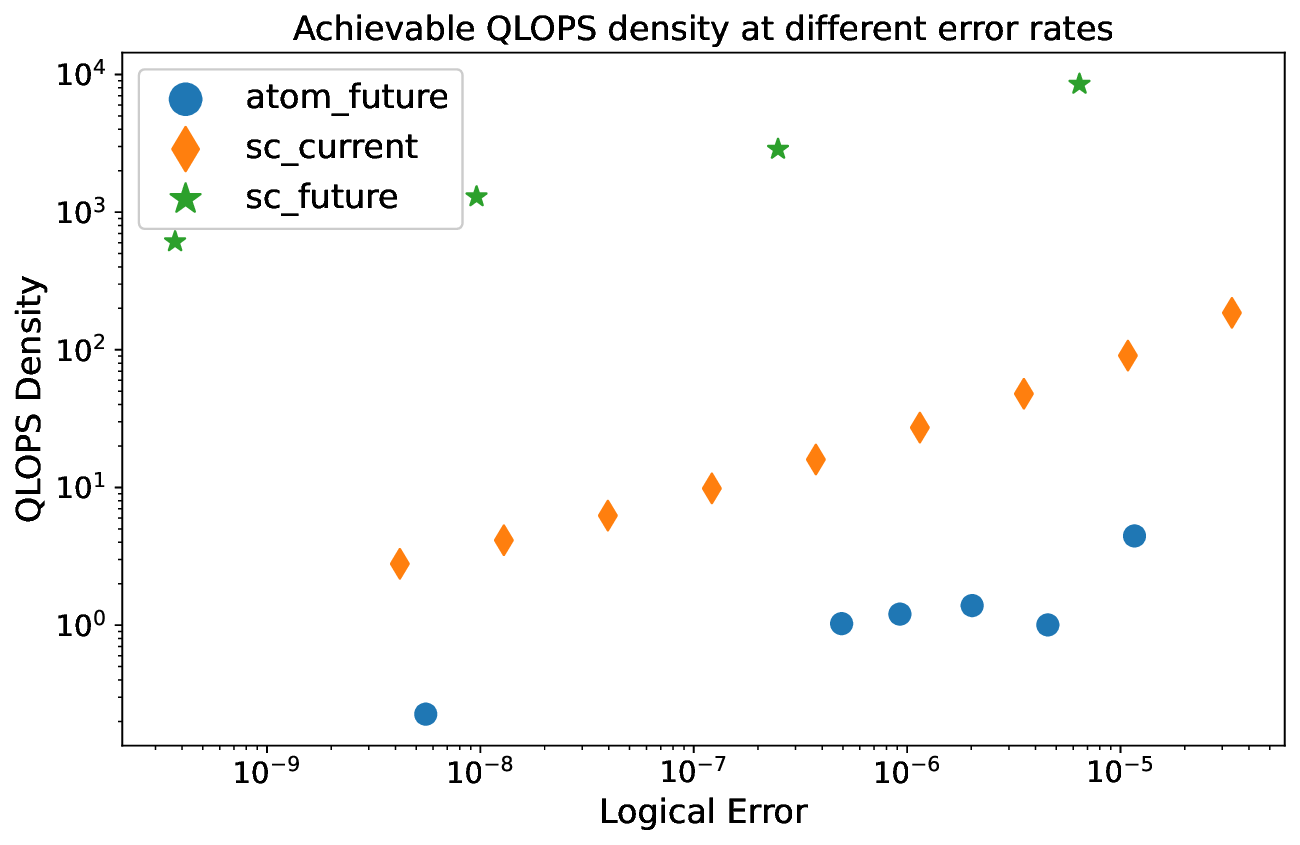}
    \caption{The achievable values of QLOPS density at various error rates for future neutral atom platform, as well as current and future superconducting platform. We did not list the current neutral atom platform because it is above the error correction threshold of GB codes. We can see that the [[72, 12, 6]] code with all syndromes is not a good scheme, because [[90, 8, 10]] with only Z syndromes has lower error rate and higher value of QLOPS density. We would like to emphasize that although superconducting qubits exhibit a higher QLOPS value than neutral atoms with the same number of qubits, this does not signify the superiority of the former platform. As some might anticipate, atomic systems are likely to scale up more readily than their superconducting counterparts.}
    \Description{The achievable values of QLOPS density at various error rates for future neutral atom platform, as well as current and future superconducting platform.}
    \label{fig:qlops}
\end{figure}

Based on the data in Table~\ref{tab:performance} we can calculate the QLOPS of the device. To make a comparison with the superconducting hardware with the same logical error rate level, we use the values of $p_L$ to find $p_0$ defined in Eq.~\eqref{eq:p0}, the logical error rate per layer per logical qubit, and find the surface code distance needed to reach this error level. Note that this can be slightly unfair for the surface code because the distance is rounded up. Then we can calculate the QLOPS of the superconducting device and make a comparison. The result can be found in Table~\ref{tab:qlops} and Fig.~\ref{fig:qlops}. 

\begin{table*}[ht]
\centering
\caption{QLOPS of GB codes and the comparison with surface codes at the same logical error rate. In the final few columns, each cell is divided into three parts, where the upper part describes future neutral atom platform and the following two parts describe the superconducting system with matching logical error rate, evaluated with current and future physical parameters. We did not list the current neutral atom platform because it is above the error correction threshold of GB codes. Besides the values of QLOPS, we also list the ``QLOPS density'' for comparison, which is defined as QLOPS divided by $N$, the total number of physical qubits.}
\label{tab:qlops}

\begin{tabular}{c|c|c|c|c|c}
\hline
Code & $p_0$ & Distance & \shortstack{Physical qubit \\ number} & QLOPS & QLOPS density \\
\hline
\multirow{3}{*}{[[72, 12, 6]] Z} & \multirow{3}{*}{1.1633e-05} & 6 & 144 & 640.35 & 4.4468 \\ \cline{3-6}
& & 13 & 4044 & 367197.06 & 90.8005 \\ \cline{3-6}
& & 5 & 588 & 5000000.00 & 8503.4014 \\ \hline
\multirow{3}{*}{[[90, 8, 10]] Z} & \multirow{3}{*}{2.0177e-06} & 10 & 180 & 249.80 & 1.3878 \\ \cline{3-6}
& & 17 & 4616 & 125707.10 & 27.2329 \\ \cline{3-6}
& & 7 & 776 & 2222222.22 & 2863.6884 \\ \hline
\multirow{3}{*}{[[108, 8, 10]] Z} & \multirow{3}{*}{9.2503e-07} & 10 & 216 & 259.85 & 1.2030 \\ \cline{3-6}
& & 19 & 5768 & 92102.23 & 15.9678 \\ \cline{3-6}
& & 7 & 776 & 2222222.22 & 2863.6884 \\ \hline
\multirow{3}{*}{[[144, 12, 12]] Z} & \multirow{3}{*}{4.9307e-07} & 12 & 288 & 295.45 & 1.0259 \\ \cline{3-6}
& & 19 & 8652 & 138153.35 & 15.9678 \\ \cline{3-6}
& & 7 & 1164 & 3333333.33 & 2863.6884 \\ \hline
\multirow{3}{*}{[[288, 12, 18]] Z} & \multirow{3}{*}{5.5451e-09} & 18 & 576 & 130.01 & 0.2257 \\ \cline{3-6}
& & 27 & 17484 & 48788.42 & 2.7905 \\ \cline{3-6}
& & 11 & 2892 & 1764705.88 & 610.2026 \\ \hline
\multirow{3}{*}{[[72, 12, 6]] ALL} & \multirow{3}{*}{4.5646e-06} & 6 & 144 & 144.59 & 1.0041 \\ \cline{3-6}
& & 15 & 5388 & 258397.93 & 47.9580 \\ \cline{3-6}
& & 7 & 1164 & 3333333.33 & 2863.6884 \\ \hline
\end{tabular}

\end{table*}

\subsection{Observations}
Several observations can be made based on the data we have obtained.
\begin{enumerate}
    \item To encode the same number of logical qubits, surface codes require around 30 times more physical qubits than the GB codes. However, the neutral atom platform still has much lower QLOPS and QLOPS density, due to the significantly longer syndrome extraction cycle length. Still, if we have the same number of qubits on both platforms, neutral atom hardware can encode much more logical qubits and run quantum applications beyond the capability of superconducting hardware. Given that the neutral atom platform might be easier to scale up \cite{Schlosser2023scalable, Pause2024supercharged, manetsch2024tweezer}, it can have even more advantage in this aspect.
    \item Taking into account $X$ syndromes in addition to $Z$ syndromes can indeed improve the decoding accuracy, but the decoding becomes much slower. Unless a much more efficient decoder can be found (Relay-BP \cite{mller2025improved} might be a candidate), this might not be worth it in terms of the QLOPS benchmark.
    \item Even though BP-based decoders are generally believed to be much slower compared to matching algorithms, they do not form a bottleneck on the neutral atom platform, as the decoding time is comparable to the time for $d$ rounds of syndrome extraction.
\end{enumerate}

\subsection{Cost of magic state distillation}
\label{subsec:distillation}
The discussions above only involve the computational qubits, but in practical quantum computation, magic state distillation constitutes a much higher overhead. In this subsection, we discuss this cost based on the scheme proposed in \cite{Litinski2019magicstate} as an example. It would be interesting to see how some of the more resource-efficient schemes, for example magic state cultivation \cite{gidney2024magic}, could be adapted to the neutral atom platform.

With the framework given in \cite{Litinski2019magicstate}, we first numerically evaluate the logical error rate per round of the surface code on each platform, and then replace Eq.~(7) in \cite{Litinski2019magicstate} by our fitted result to study the overall infidelity of the distillation process. We need to choose the parameters of the distillation schemes so that the infidelity of the output magic states is smaller or equal to the logical error rate. Then we can estimate the number of qubits needed to produce a magic state for each logical operation. Note that for neutral atoms, the lengths of the syndrome extraction cycle are different for the computation and distillation, as they are encoded with different codes.

We have searched through the one-level 15-to-1 scheme with all possible parameters in a reasonable range. For two-level schemes, an exhaustive search is not possible due to the large number of parameters. We tried some choices of the parameters, but found that the result is suboptimal compared to the one-level 15-to-1 scheme, due to their large number of qubits and cycles needed for the distillation.

The distillation protocol and its cost can be found in Table~\ref{tab:magic}. In the final column, we list the total number of qubits used for distillation, so that a magic state can be produced for each logical qubit in each logical cycle on average. This is given by the number of qubits in each distillation unit multiplied by the number of distillation units needed. Note that the calculation of the number of distillation units takes into account the syndrome extraction cycle difference and the post-selection rate, and is rounded up to an integer.

\begin{table*}
    \centering
    \caption{The cost of magic state distillation for each choice of GB code. We choose the parameters for the protocols in \cite{Litinski2019magicstate} so that the error rate matches the logical error rate. We list the the distillation error rate, number of qubits in the distillation factory and the number of syndrome extraction cycles needed by the protocol (including the cost of post-selection). In the final column we list the total number of qubits used for distillation so that one magic state can be produced for each logical qubit in each logical cycle on average.}
    \label{tab:magic}
    \footnotesize
    \begin{tabular}{c|c|c|c|c|c|c}
        \hline
        Code & $p_0$ & Protocol & Distillation error & \shortstack{Qubits per \\ distillation unit} & \shortstack{Number of \\ cycles} & \shortstack{Total number \\ of qubits} \\
        \hline
        \multirow{3}{*}{[[72, 12, 6]] Z} & \multirow{3}{*}{1.1633e-05} & (15-to-1)${}_{9,3,3}$ & 2.3317e-06 & 1146 & 18.6423 & 2292 \\ \cline{3-7}
        & & (15-to-1)${}_{23,9,9}$ & 9.0607e-06 & 8178 & 58.0795 & 155382 \\ \cline{3-7}
        & & (15-to-1)${}_{7,3,3}$ & 1.1269e-05 & 810 & 18.6403 & 30780 \\ \hline
        \multirow{3}{*}{[[90, 8, 10]] Z} & \multirow{3}{*}{2.0177e-06} & (15-to-1)${}_{11,3,3}$ & 9.5818e-07 & 1530 & 18.7980 & 1530 \\ \cline{3-7}
        & & (15-to-1)${}_{29,9,11}$ & 1.9588e-06 & 11354 & 70.1846 & 90832 \\ \cline{3-7}
        & & (15-to-1)${}_{9,3,3}$ & 1.3258e-06 & 1146 & 18.8261 & 19482 \\ \hline
        \multirow{3}{*}{[[108, 8, 10]] Z} & \multirow{3}{*}{9.2503e-07} & (15-to-1)${}_{11,5,3}$ & 6.8826e-07 & 2058 & 18.6779 & 2058  \\ \cline{3-7}
        & & (15-to-1)${}_{29,11,11}$ & 5.5174e-07 & 12746 & 68.1534 & 76476 \\ \cline{3-7}
        & & (15-to-1)${}_{11,3,5}$ & 9.8766e-08 & 1538 & 31.1579 & 43064 \\ \hline
        \multirow{3}{*}{[[144, 12, 12]] Z} & \multirow{3}{*}{4.9307e-07} & (15-to-1)${}_{11,3,5}$ & 2.5036e-07 & 1538 & 30.9210 & 3076 \\ \cline{3-7}
        & & (15-to-1)${}_{31,11,11}$ & 4.7231e-07 & 13994 & 68.2899 & 125946 \\ \cline{3-7}
        & & (15-to-1)${}_{11,3,5}$ & 9.8766e-08 & 1538 & 31.1579 & 64596\\ \hline
        \multirow{3}{*}{[[288, 12, 18]] Z} & \multirow{3}{*}{5.5451e-09} & (15-to-1)${}_{15,5,5}$ & 2.1524e-09 & 3170 & 30.1425 & 3170 \\ \cline{3-7}
        & & (15-to-1)${}_{39,17,15}$ & 4.4065e-09 & 25098 & 90.6157 & 100392 \\ \cline{3-7}
        & & (15-to-1)${}_{13,5,5}$ & 1.3677e-09 & 2594 & 30.0962 & 57068 \\ \hline
        \multirow{3}{*}{[[72, 12, 6]] ALL} & \multirow{3}{*}{4.5646e-06} & (15-to-1)${}_{9,3,3}$ & 2.3317e-06 & 1146 & 18.6423 & 1146 \\ \cline{3-7}
        & & (15-to-1)${}_{25,11,9}$ & 4.3460e-06 & 10386 &  57.4921 & 135018 \\ \cline{3-7}
        & & (15-to-1)${}_{9,3,3}$ & 1.3258e-06 & 1146 & 18.8261 & 29796 \\ \hline
    \end{tabular}
\end{table*}

\subsection{Relation to recent works on RSA2048 resource analysis}
Recently, two papers \cite{gidney2025factor,zhou2025resource} have been posted to analyze the resources required to factor a 2048-bit RSA integer on the superconducting and neutral atom platforms, respectively. In this subsection, we benchmark the devices in these papers using QLOPS. Although QLOPS is an application-independent benchmark and might not fully capture the performance for specific tasks, we find that its estimation roughly matches the results provided in the papers in some aspects.

For the superconducting system in \cite{gidney2025factor}, there are 1280+131 computational logical qubits (excluding the qubits associated with magic state distillation). Note that the 1280 logical qubits are in cold storage and do not actively engage in any logical operations, but they are still taken into account for the QLOPS computation. The active qubits are encoded with $d=25$ surface codes, with a syndrome extraction cycle of \us{1} and a reaction time of \us{10}. With Eq.~\eqref{eq:qlops} we find that the QLOPS value is $Q_\SC = 4.0314\times 10^7$. The total number of data qubits is $1280 \times 430 + 131 \times (2d^2-1)=714019$, and QLOPS divided by the number of data qubits equals $56.4611$.

For the neutral atom system in \cite{zhou2025resource}, there are a total of 19 million physical qubits with $d=27$ surface code encoding. Each code patch contains $d^2$ data qubits and $d^2-1$ ancilla qubits. There are 192 magic state distillation factories, each containing $3\times 12$ code patches. Excluding the qubits in the factories, we find that there are 6128 computational logical qubits. The syndrome extraction cycle is \us{900} and the reaction time is \us{1000}. Correlated decoding has been adopted in the system, which allows $O(1)$ syndrome extraction cycles per transversal logical operation. The reaction time still has some influence on the overall performance, but in a complicated application-specific way shown in their Fig.~7. For the QLOPS estimation, we simply ignore the reaction time and modify Eq.~\eqref{eq:qlops} into $Q=k\times 1/t_\SEC$ (assuming 1 syndrome extraction cycle following each transversal gate according their Section~IV.2). The value of QLOPS is $Q_\atom=6.8089\times 10^6$ and QLOPS divided by the total number of data qubits is 0.7626.

Multiplying QLOPS with the running time leads to an estimation of the total amount of logical operations of the quantum algorithm. With the estimated running time of $t_\SC = 4.96$ days and $t_\atom = 5.6$ days in the two papers, we find that
\begin{equation}
    \frac{Q_\SC t_\SC}{Q_\atom t_\atom} = 5.244,
\end{equation}
which deviates from 1 by a factor of 5. This deviation is caused by the discrepancy between QLOPS as a general-purpose benchmark and the application-specific scenario, as well as the difference of the algorithms used due to their choices of parameters.

If we estimate the total number of operations in the algorithm based on the Toffoli count, there are $n^T_\SC = 6.5\times 10^9$ for superconducting qubits (according to \cite[Table~5]{gidney2025factor}) and $n^T_\atom = 3\times 10^9$ Toffoli gates for neutral atoms (according to \cite[Section~III.6]{zhou2025resource}), so the deviation becomes
\begin{equation}
    \frac{Q_\SC t_\SC / n^T_\SC}{Q_\atom t_\atom / n^T_\atom} = 2.4204. \label{eq:compare}
\end{equation}

$n^T/Q$ provides a lower bound for the running time of a quantum algorithm on a specific hardware platform, as $Q$ describes the maximum number of logical operations on the device by assuming maximal parallelization of the gates, and $n^T$ is a lower bound of the total number of logical gates. A slightly more careful estimation can be obtained by noticing that the implementation of a Toffoli gate by gate teleportation requires $\sim 10$ Clifford gates \cite{gidney2019flexible}. Both \cite{zhou2025resource} and \cite{gidney2025factor} used slightly improved designs by combining some of the gates with other logical gates, but 10 would still be a good estimation. Still, $10 n^T/Q$ is an underestimation of the total running time, as some logical qubits are idling and the logical Clifford gates in the circuit are not taken into account. $t/(10n^T/Q)$ describes the ratio of underestimation and is $\sim 110$ for neutral atoms and $\sim 270$ for superconducting qubits. This leaves space for optimization by, e.g., better compilation and improved routing so that more Toffoli gates can be applied in each logical cycle.

Although $t$ can be 2 orders larger than $10n^T / Q$ in the RSA2048 analysis, indicating that this lower bound is loose for this specific task, from Eq.~\eqref{eq:compare} we can see that the ratio of overestimation is similar for the two platforms. This means that QLOPS is still a great indicator for comparing different hardware platforms.

\subsection{Interpretation of QLOPS}
As explained earlier, FLOPS measures a computer's performance by quantifying how many floating-point operations it can execute per second. Similarly, we defined QLOPS as an analogous metric to assess theoretical performance using quantum hardware parameters. While QLOPS effectively estimates system performance, there are subtle but important differences between QLOPS and FLOPS that we should note.

Although real-world performance depends on many factors like memory bandwidth and parallel efficiency, it often achieves a significant fraction of the theoretical FLOPS. Thus, FLOPS serves as a key metric for comparing computational power, especially in scientific computing, machine learning, and high-performance computing (HPC). Higher FLOPS values indicate greater processing capability, with modern GPUs and supercomputers often rated in teraflops ($10^{12}$ FLOPS), petaflops ($10^{15}$ FLOPS), or beyond. 

Like FLOPS, QLOPS is designed to measure overall system performance by accounting for multiple contributing factors. As discussed above, while QLOPS shares connections with recent RSA2048 resource estimations, it is not application-specific. Consequently, it offers limited insights for particular applications---which is unsurprising given its general nature. Nevertheless, the interesting observation is that, in the superconducting system \cite{gidney2025factor} there are only 6 magic state distillation factories, and their combined production rate is around 1 Toffoli state per logical cycle. This means that only 1 Toffoli gate is applied per logical cycle on average, or 10 logical gates of gate teleportation. This partially explains the $10^2$ factor of discrepancy, since each of the $\sim 10^3$ logical qubits is expected to perform a logical operation per logical cycle in the QLOPS calculation, while only 10 are actually applied in the circuit. Similarly, in the neutral atom system \cite{zhou2025resource}, there are 192 factories that each require $\sim 100$ SECs to produce a Toffoli state, so on average there are 1 or 2 Toffoli states produced per SEC, which translates to 10 to 20 logical gates in the $10n^T$ estimation. There are $\sim 6000$ logical qubits, and each is expected to perform a logical operation per SEC, so this limitation posed by magic state production also partially explains the $10^2$ discrepancy. Another interpretation of the discrepancy is that there are Clifford gates in the circuits in addition to Toffoli gates, which has not been taken into account. This is a consequence of the lack of knowledge about the algorithm, not the drawback of the definition of QLOPS. If the total number of gates in the circuit instead of just Toffoli gates is known, one might get a more accurate estimation.

This discrepancy arises primarily from the limited parallelization capability of logical Toffoli operations in current fault-tolerant quantum architectures. While this constraint highlights potential optimization opportunities, it also raises a fundamental question: Given the above-mentioned practical challenges, should QLOPS---as a \emph{theoretical upper bound for quantum hardware performance}---be revisited or refined in its formulation?

\section{Discussion}
\label{sec:discussion}
In this work, we proposed QLOPS as a benchmark for fault-tolerant quantum computing hardware. We presented the framework of analyzing the QLOPS of a quantum system, and as a demonstration we calculated the QLOPS of the superconducting and neutral atom systems. We would like to emphasize that this calculation does not represent the full potential of these hardware platforms: we have only picked one specific fault-tolerant scheme for each of the systems as a demonstration, while leaving the optimization of the QLOPS value for future work. As an indicator of quantum hardware performance, QLOPS can provide guidance on improving the capability of a system. One can tweak various aspects of the system and see how the QLOPS changes, and some examples include:
\begin{itemize}
    \item The physical noise parameters;
    \item The quantum error-correction code and the associated FTQC schemes that are consistent with the system;
    \item The decoder.
\end{itemize}

While QLOPS characterizes the computational capability of a quantum hardware by capturing its key aspects, there are also other factors that might impact the hardware performance, but have not been taken into account. Some examples include:
\begin{enumerate}
    \item The transmission latency of syndromes to the decoder;
    \item The bandwidth constraints of syndrome data transmission;
    \item The efficiency loss due to the use of a parallel window decoder;
    \item The auxiliary qubits used for logical operations, e.g., the routing space for lattice surgery, and the adapters between the GB codes and surface codes introduced in \cite{viszlai2023matching}.
\end{enumerate}
We leave it to future work to include these factors in a framework of benchmarking fault-tolerant quantum computing hardware.

\begin{acks}
This work is supported by Zhongguancun Laboratory and the National Key Research and Development Program of China (Grant No. 2025YFE0200900). We thank Xiaobo Zhu, Yulin Wu, Fei Yan, Chunqing Deng, Yingfei Gu, and Wenlan Chen for sharing their insights on future quantum hardware parameters. The data has been anonymized and aggregated from multiple sources to enhance privacy protection while ensuring robustness.

\end{acks}

\bibliographystyle{ACM-Reference-Format}
\bibliography{qlops-acm}


\end{document}